\documentclass[aip,jap,preprint,amsmath,amssymb]{revtex4-1}
\usepackage{amsmath}
\usepackage{amssymb}
\usepackage{graphicx}
\usepackage{hyperref}
\begin{document}
\title{Effects of n-type doping in InAs/GaAs quantum dot layer on current-voltage characteristic of intermediate band solar cells}
\author{Yong-Xian Gu}
\author{Xiao-Guang Yang}
\author{Hai-Ming Ji}
\author{Peng-Fei Xu}
\author{Tao Yang} 
\email{tyang@semi.ac.cn}
\affiliation{Key Laboratory of Semiconductor Materials, Institute of Semiconductors, Chinese Academy of Sciences, P.O. Box 912, Beijing 100083, China}

\begin{abstract}
We investigated the current-voltage characteristic of InAs/GaAs quantum dot intermediate band solar cells (QD IBSCs) with different n-type doping density in the QD layer. The n-type doping evidently increases the open circuit voltage, meanwhile decreases the short circuit current density, and leads to the conversion efficiency approaching that of the control solar cell, that is the major role of n-type doping is to suppress the effects of QDs on the current-voltage characteristic. Our model adopts practical parameters for simulation rather than those from detailed balanced method, so that the results in our simulation are not overestimated.
\end{abstract}
\maketitle

\section{Introduction}
Intermediate band solar cells (IBSCs) are expected to be the third generation photovoltaic devices because of the potential for high energy conversion efficiency.\cite{luque_increasing_1997, luque_thermodynamic_2001, nozawa_detailed_2011} In addition to the absorption between the conduction band (CB) and the valence band (VB) the same as single junction solar cells, there are transitions from the VB to the IB and from the IB to the CB, it is expected that the IBSCs have the advantages of large increase of short circuit current density, small decrease of open circuit voltage, and enhanced conversion efficiency. 

Quantum dots (QDs) are proposed to form the IB,\cite{laghumavarapu_gasbgaas_2007, popescu_theoretical_2008, guimard_fabrication_2010, bailey_near_2011} and another implementation is employing alloys with intrinsic IB.\cite{lopez_engineering_2011, luque_photovoltaics:_2011} By now the experiments show that the efficiency of InAs/GaAs QD IBSCs are lower than control GaAs solar cells without QDs due to the reduction of open circuit voltage and small increase of short circuit current density.\cite{blokhin_algaas/gaas_2009, guimard_fabrication_2010, bailey_near_2011}

N-type doping in the i-region are raised to half-fill the IB so that the IB can provide strong absorption in transitions from the IB to both the VB and the CB.\cite{marti_partial_2001} Modulated $\delta$-doping is generally carried out with sheet density equal to the QD density,\cite{marti_novel_2006, marti_production_2006, lin_model_2009} the doped layer is a certain distance away from the QD layer, and there is approximately one electron on the dot level per quantum dot because there are two states with different spin. Recently direct doping Si into QDs not only can half-fill the IB but also lead to enhancement of the photoluminescence intensity by a dedicated design doping in the QD's assembling stage,\cite{inoue_impurity_2010} then this method is applied to the IBSC and increase in photocurrent is observed.\cite{okada_increase_2011}

In this paper, we employ the drift-diffusion model to investigate the effects of InAs/GaAs QDs on current-voltage characteristic in IBSCs by changing n-type doping density in the QD layer. The main difference between this modeling and previous work done on IBSCs, is that previously the current-voltage characteristics are mostly based on the detailed balanced method,\cite{luque_increasing_1997, navruz_efficiency_2008, navruz_detailed_2009, nozawa_detailed_2011} material parameters such as the absorption coefficients, carrier mobilities, etc. are not involved. Although some works adopt drift-diffusion model for the simulation of the device structure, the generation and recombination via the IB are still deduced from the detailed balanced method,\cite{lin_model_2009, strandberg2011} because of the large difference between the theoretical and experimental results, it is necessary to adopt practical parameters.

\section{The theory model}
\subsection{Drift-diffusion model}
In solar cells, the electron and hole current equations include both drift and diffusion components,
\begin{equation}
J_n=en\mu_n\frac{d\psi}{dx}-eD_n\frac{dn}{dx},   \\
\end{equation}
\begin{equation}
J_p=-ep\mu_p\frac{d\psi}{dx}-eD_p\frac{dp}{dx},
\end{equation}
where $e$ is the electron charge, $n$ and $p$ are the electron and hole concentration, $\mu_n$ and $\mu_p$ are the carrier mobilities, which decrease to small values as doping density increases, $D_n$ and $D_p$ are the carrier diffusion constants determined by the Einstein relation, $D_n=\frac{kT}{e}u_n$ and $D_p=\frac{kT}{e}u_p$.

Drift-diffusion model includes Poisson equation, electron and hole current continuity equations,\cite{luque_operation_2006,lin_drift-diffusion_2009,strandberg2011} when considering the role of QDs, the three equations can be written as
\begin{eqnarray}
 -\frac{d}{dx}\left(\varepsilon\frac{d\psi}{dx}\right)&=e\left[p-n-fN_I+N_D^+-N_A^-\right], \label{eq:poisson} \\
  \frac{dJ_n}{dx}&=e(G_{CV}+G_I-R_{CV}-R_{SRH}), \label{eq:jn} \\
  \frac{dJ_p}{dx}&=e(G_{CV}+G_I-R_{CV}-R_{SRH}), \label{eq:jp}
\end{eqnarray}
where $\varepsilon$ is the permittivity of the medium, $\psi$ is the electrostatic potential, $N_I$ is the number of the IB states per unit volume, $f$ is the electron occupation factor in the IB, which will be studied later, $N_D^+$ is the ionized donors and $N_A^-$ is the ionized acceptors.  

$G_{CV}=\int \alpha(\lambda)I_0(\lambda)R(\lambda)\exp[-\alpha(\lambda)x]d\lambda$, is the generation rate of carriers from the VB to CB, where $I_0(\lambda)$ is the intensity of the light at $x=0$, $\alpha(\lambda)$ is the absorption coefficient of the host material, and $R(\lambda)$ is the reflectivity of the front surface. 

$G_I$ is the generation rate of electron hole pairs through absorption of the IB, which will be studied later. $R_{CV}=r_{CV}\left(np-n_i^2\right)$ is the direct recombination between the CB and the VB, and $r_{CV}$ is the recombination coefficient.

$R_{SRH}$ is the Shockley-Read-Hall (SRH) recombination between conduction and valance bands,
\begin{equation}
\label{eq:srh}
R_{SRH}=\frac{\left(np-n_i^2\right)}{\tau_n(p+p_1)+\tau_p(n+n_1)},
\end{equation}
where $\tau_n$ and $\tau_p$ are electron and hole lifetime respectively, $n_1=n_i\exp\left(\frac{E_t-E_i}{kT}\right)$ and $p_1=n_i\exp\left(\frac{E_i-E_t}{kT}\right)$, $n_i$ is the intrinsic carrier concentration of the host material, $E_t$ is the defect energy position, which for simplicity is set in the middle of GaAs bandgap, $E_i$ is the intrinsic Fermi energy, $k$ is the Boltzmann constant and $T$ is the temperature of the solar cell.

\subsection{Electron occupation factor and net generation}
In most literatures, carrier generation and recombination via the IB are analyzed based on detailed balanced method, finally the results are often overestimated and deviate from experimental results. By adding the light generation components to the derivation process of SRH recombination in textbook,\cite{Sze_2007} the involved four processes for the generation and recombination can be written as:\cite{luque_operation_2006} The carrier recombination from the CB to the IB
\begin{equation}                                                                                                                                       
 R_{CI}=r_{CI}N_{I}(1-f)n,
\end{equation}
the generation from the IB to the CB 
\begin{equation}
 G_{CI}=e_{e}N_{I}f+g_{CI}N_{I}f,
\end{equation}
the recombination from the IB to the VB
\begin{equation} 
 R_{VI}=r_{VI}N_{I}fp,
\end{equation}
and the generation from the VB to the IB 
\begin{equation} 
 G_{VI}=e_{h}N_{I}(1-f)+g_{VI}N_{I}(1-f),
\end{equation}
where $r_{CI}$ and $r_{VI}$ are the recombination coefficients, $e_e$ and $e_h$ are the emitting coefficients due to factors such as thermal excitation, $g_{CI}$ and $g_{VI}$ are light generation coefficients proportional to the light intensity and absorption coefficient via the IB.

If there are no optical and electrical injection, the device is in equilibrium with a uniform Fermi level $E_{f0}$, electron density $n_0$ in the CB, hole density $p_0$ in the VB and electron occupation $f_0$ in the IB, and recombination is equal to generation,
\begin{equation}
r_{CI}N_{I}(1-f_0)n_0=e_{e}N_{I}f_0,
\end{equation}
\begin{equation}
r_{VI}N_{I}f_0p_0=e_{h}N_{I}(1-f_0).
\end{equation}
Using 
\begin{equation}
n_0=n_i\exp\left(\frac{E_{f0}-E_i}{kT}\right),
\end{equation}
\begin{equation}
p_0=n_i\exp\left(\frac{E_i-E_{f0}}{kT}\right),
\end{equation}
and 
\begin{equation}
f_0=\frac{1}{\exp\left(\frac{E_{I}-E_{f0}}{kT}\right)+1},\label{eq:f0}
\end{equation}
where $E_{I}$ is the IB energy position, we can get $e_e$ and $e_h$,
\begin{equation}\label{eq:ee}
e_e=r_{CI}n_i\exp{\left(\frac{E_I-E_i}{kT}\right)},
\end{equation}
\begin{equation}
e_h=r_{VI}n_i\exp{\left(\frac{E_i-E_I}{kT}\right)}.
\end{equation}

If there is steady light injection, electrons have a stable transition from the VB to the CB via the IB with net transition rate $G_I$,
\begin{equation} \label{eq:gi_1}
G_{I}\equiv G_{CI}-R_{CI}=G_{VI}-R_{VI}.
\end{equation}
Solve the Eq.~(\ref{eq:gi_1}), leading to \cite{luque_operation_2006}
\begin{equation}
\label{eq:f}
f=\frac{e_h+g_{VI}+r_{CI}n}{e_e+e_h+g_{CI}+g_{VI}+r_{CI}n+r_{VI}p},
\end{equation}
\begin{equation}
\label{eq:gi}
G_I=\frac{N_I\left[e_eg_{VI}+e_hg_{CI}+g_{CI}g_{VI}-r_{CI}r_{VI}(pn-n_i^2)\right]}{e_e+e_h+g_{CI}+g_{VI}+r_{CI}n+r_{VI}p}.
\end{equation}
This time if we set $g_{CI}=g_{VI}=0$, $f$ and $G_I$ can be expressed as
\begin{equation}
f^{'}=\frac{e_h+r_{CI}n}{e_e+e_h+r_{CI}n+r_{VI}p}, \label{eq:fp}
\end{equation}
\begin{equation}
G_{I}^{'}=-\frac{N_Ir_{CI}r_{VI}(pn-n_i^2)}{e_e+e_h+r_{CI}n+r_{VI}p}. \label{eq:gip}
\end{equation}
$G_{I}$ has the form of SRH recombination, if there is no electrical injection, $n=n_0$ and $p=p_0$, so Eq.~\ref{eq:fp} becomes a equivalent expression for Eq.~\ref{eq:f0},
\begin{equation}
f_0=\frac{1}{\exp\left(\frac{E_{I}-E_{f0}}{kT}\right)+1}=\frac{e_h+r_{CI}n_0}{e_e+e_h+r_{CI}n_0+r_{VI}p_0}.\label{eq:f0_2}
\end{equation}

The selective ohmic contact boundary conditions for Eqs.~(\ref{eq:jn}) and (\ref{eq:jp}) are used when solving the drift-diffusion model,\cite{fonash_solar_2010} $J_n(L)=eS_{nL}[n(L)-n_0(L)]$, $J_p(L)=-eS_{pL}[p(L)-p_0(L)]$, $J_n(R)=-eS_{nR}[n(R)-n_0(R)]$ and $J_p(R)=eS_{pR}[p(R)-p_0(R)]$, $L$ ($R$) stands for left (right) boundary, $S$s with subscripts are surface recombination coefficients. 

\section{Simulation and results}
\subsection{Parameters}
\begin{figure}[ht]
 \centering
 \includegraphics[width=0.9\columnwidth,clip]{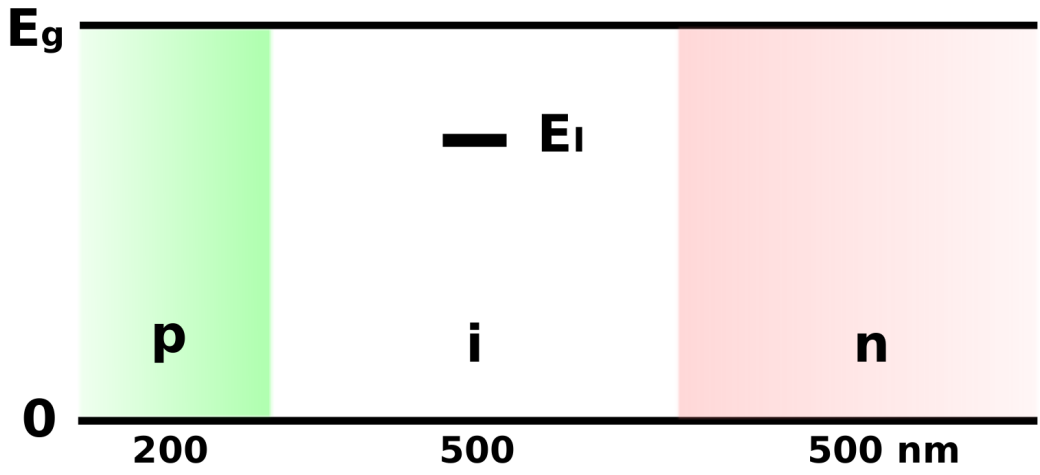}
 \caption{(Color online) The simulation structure, the 50~nm QD layer is placed in the middle of the i-region.}
 \label{fig:fig1}
\end{figure}

The simulation structure is depicted in Fig.~\ref{fig:fig1}, the thickness of the p-region, i-region and n-region are 200, 500 and 500 nm respectively. The 50 nm QD layer is placed in the center of the i-region. The bandgap of the host material GaAs $E_g$ is $1.424$~eV, the IB energy level $E_I$ is formed by electron ground energy level of QD, for InAs/GaAs QD $E_I=1.124$~eV.\cite{luque_operation_2006} The solar cell is simulated under AM~1.5 \cite{fonash_solar_2010} solar spectrum with the reflectivity of the front surface $R(\lambda)$ setting to 0.1, and GaAs absorption coefficient from Ref.~\onlinecite{casey_1974}, the carrier mobilities of GaAs are related to the doping,
\begin{equation}
\mu_n=1000+\frac{7200}{1+[(N_D+N_A)/6\times10^{16}]^{0.55}}~\mathrm{cm^2/Vs},  \label{eq:mun}
\end{equation}
\begin{equation}
\mu_p=32+\frac{400}{1+[(N_D+N_A)/1.88\times10^{17}]^{0.5}}~\mathrm{cm^2/Vs}.   \label{eq:mup}
\end{equation}

For $r_{CI}$ and $r_{VI}$, according to Eq.~\ref{eq:gip}, the lifetimes of electron and hole are $\tau_{n}'=1/(N_{I}r_{CI})$ and $\tau_{p}'=1/(N_{I}r_{VI})$, respectively. Due to Auger cooling effect,\cite{tomicacute_intermediate-band_2010} electrons are no longer affected by the phonon bottleneck effect, and can easily be captured from the CB to the IB, so $\tau_{n}'$ is small, about $1\times10^{-12}$~s, Auger cooling effect is beneficial to QD lasers for high modulation rate, but detrimental to QD solar cells. $\tau_{p}'$ is mainly determined by electron radiative recombination lifetime from the IB to the VB, $\tau_{p}'\approx1\times10^{-9}$~s. So $r_{CI}$ and $r_{VI}$ are set to $1.25\times10^{-5}~\mathrm{cm^3~s^{-1}}$ and $1.25\times10^{-8}~\mathrm{cm^3~s^{-1}}$, respectively. The light generation coefficients $g_{CI}$ and $g_{VI}$ are from Ref.~\onlinecite{luque_operation_2006}, which are from fitting the experimental results, $g_{CI}=g_{VI}=2.31\times10^3~\mathrm{s^{-1}}$, to make a comparison, we also studied the case with $g_{CI}=g_{VI}=1\times10^5~\mathrm{s^{-1}}$, which are the only parameters we assumed. Other parameters are given in Tab.~\ref{tab:parameters}.

\subsection{N-type doping without QDs}
\begin{figure}[ht]
 \centering
 \includegraphics[width=0.9\columnwidth,clip]{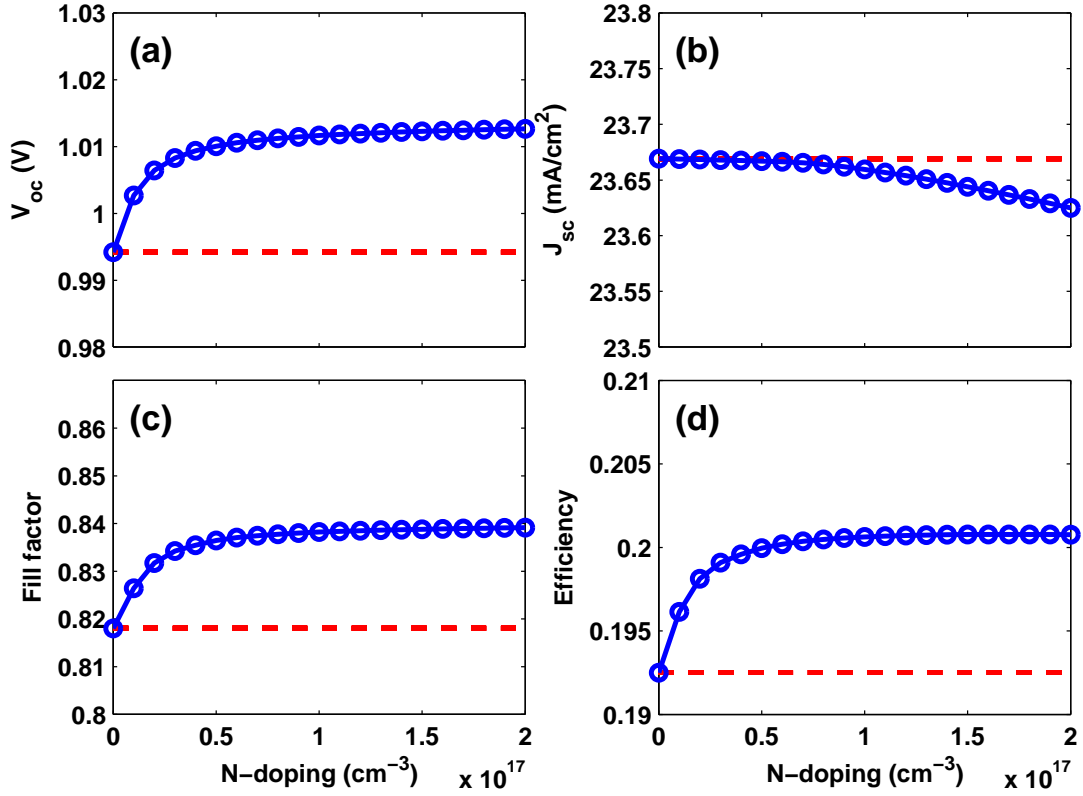}
 \caption{(Color online) Doping without QDs, the current-voltage characteristic changes as the doping density increases. (a) Open circuit voltage. (b) Short circuit current density. (c) Fill factor. (d) Conversion efficiency. Data for the control GaAs solar cell without QD layer are plotted as dashed lines.}
\label{fig:fig2}
\end{figure}

\begin{figure}[ht]
 \centering
 \includegraphics[width=0.7\columnwidth,clip]{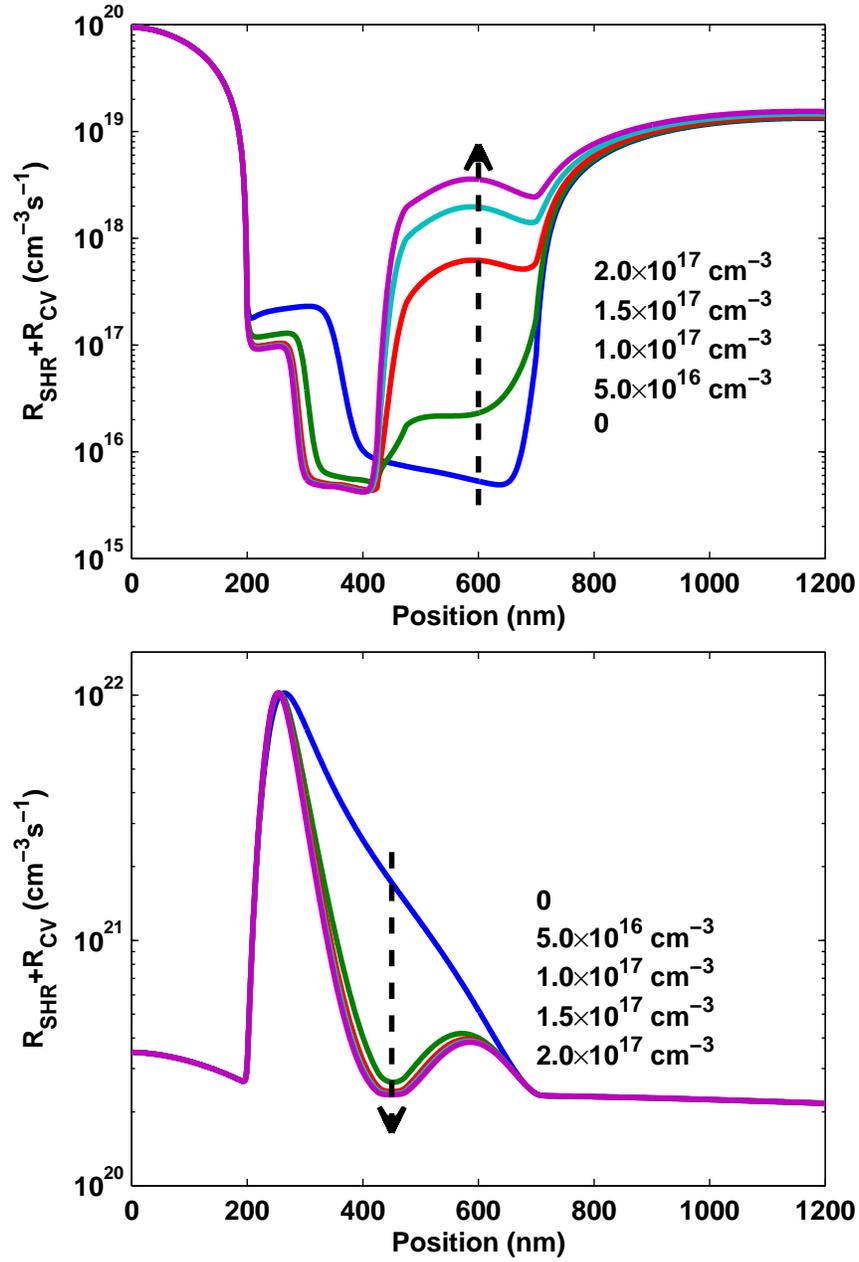}
 \caption{(Color online) Doping without QDs, the total recombination $R_{SRH}+R_{CV}$ changes as different doping density. (a) On short circuit condition. (b) When the output voltage is 1 V. Data for the control GaAs solar cell without QD layer and doping are plotted as dashed line.} 
\label{fig:fig3}
\end{figure}

In order to exclude the effect of doping on the control GaAs solar cell, we first investigate the doping in the 50 nm layer without QDs. In Fig.~\ref{fig:fig2}, as the doping density increases from 0 to $2\times10^{17}~\mathrm{cm^{-3}}$, the open circuit voltage $V_{oc}$ increases from 0.99 to 1.01 V, while the short circuit current density $J_{sc}$ decreases from 23.67 to $23.63~\mathrm{mA/cm^2}$, the fill factor from 0.82 to 0.84, and the conversion efficiency from 19.3\% to 20.1\%. When the doping density is 0, it is corresponding to the control GaAs solar cell and its data are noted as dashed line. The change of $J_{sc}$ and $V_{oc}$ are due to the change of total recombination $R_{SRH}+R_{CV}$. In Fig.~\ref{fig:fig3}(a), for short circuit the total recombination between the doping layer and n-region increases as the doping density increases, while the total recombination decreases when the output voltage is 1V as shown in Fig.~\ref{fig:fig3}(b). In fact, the doping layer has a coulomb screen effect on the region between the doping layer and the n-region, the build-in electric potential gradually falls to the region between p-region and the doping layer, and the carriers have a redistribution, so the total recombination changes accordingly. 

\subsection{Low light generation coefficients}
\begin{figure}[ht]
 \centering
 \includegraphics[width=0.9\columnwidth,clip]{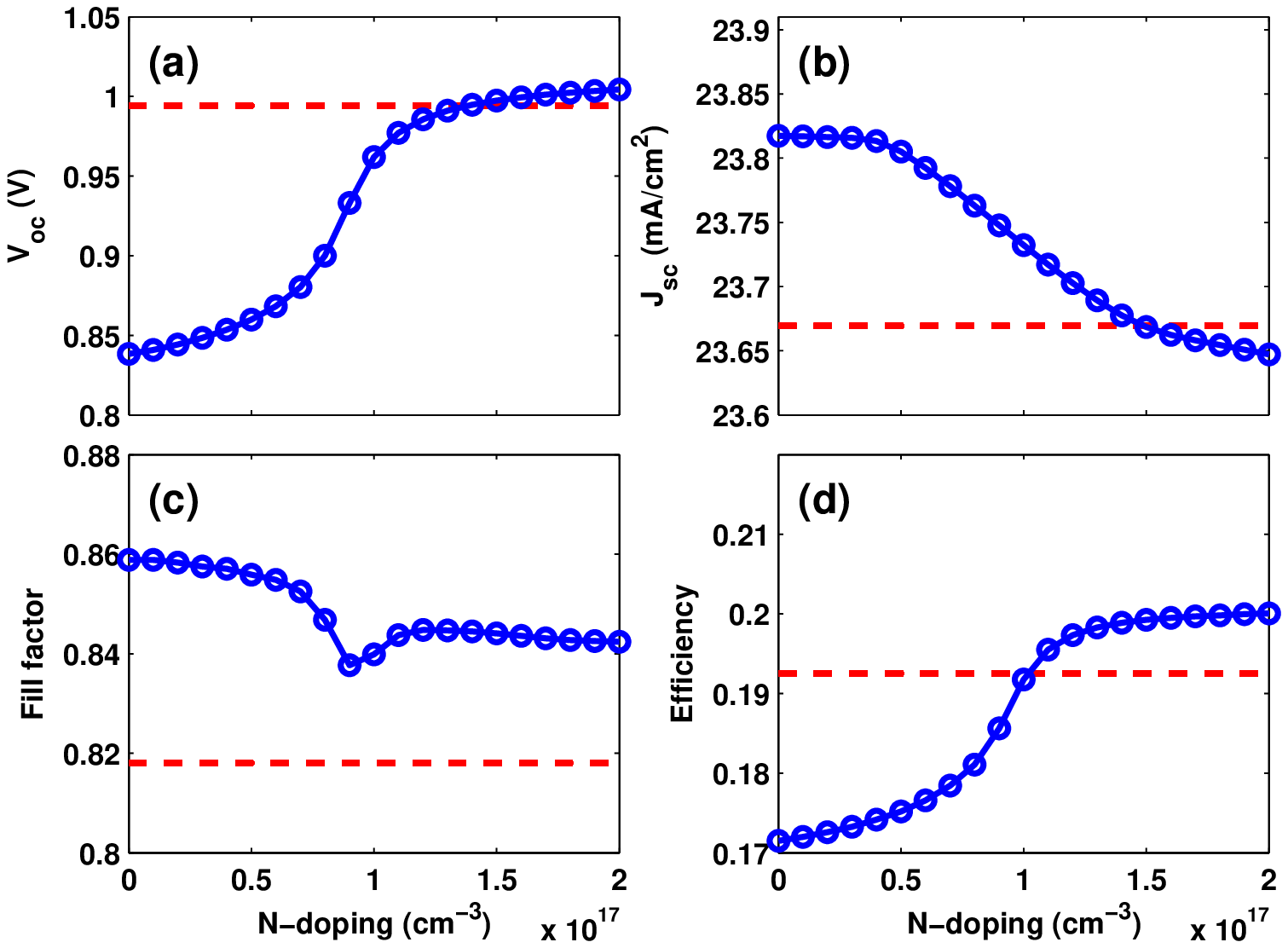}
 \caption{(Color online) For $g_{CI}=g_{VI}=2.31\times10^3~\mathrm{s^{-1}}$, the current-voltage characteristic changes as the doping density increases. (a) Open circuit voltage. (b) Short circuit current density. (c) Fill factor. (d) Conversion efficiency. Data for the control GaAs solar cell without QD layer are plotted as dashed lines.}
 \label{fig:fig4}
\end{figure}

In the calculations below, the QDs are added to the 50 nm layer. In Fig.~\ref{fig:fig4}, $g_{CI}=g_{VI}=2.31\times10^3~\mathrm{s^{-1}}$, as the doping density increases from 0 to $2\times10^{17}~\mathrm{cm^{-3}}$, $V_{oc}$ increases significantly from 0.84 to 1.0 V, for the doping density larger than $1.5\times10^{17} \mathrm{cm^{-3}}$, $V_{oc}$ is a little larger than that of the control GaAs solar cell 0.99 V. $J_{sc}$ decreases monotonically from 23.82 to 23.65~$\mathrm{mA/cm^2}$, for the control GaAs solar cell it is $23.67~\mathrm{mA/cm^2}$. The fill factor is larger than that of the control GaAs solar cell for the whole doping range, and there is a kink at doping density about $9\times10^{16}~\mathrm{cm^{-3}}$, which is a little larger than $N_I$, it indicates the kink happens when the IB is fully occupied by electrons. What we most care the conversion efficiency increases from 17.2\% to 20\%, that of the control GaAs solar cell is 19.3\%.

\subsection{High light generation coefficients}
To better understand the effects of the doping on current-voltage characteristic, we calculated another set of data for $g_{CI}$ and $g_{VI}$, $g_{CI}=g_{VI}=1\times10^5~\mathrm{s^{-1}}$, corresponding to high photon absorption or wide range absorption spectrum. The results are shown in Fig.~\ref{fig:fig5}. $V_{oc}$ has a large increase, which is almost the same to that for low generation coefficients, while $J_{sc}$ has a large increase with maximum value 30.07 $\mathrm{mA/cm^2}$, it decrease to $23.66~\mathrm{mA/cm^2}$ as the doping density increases. The fill factor has a large fluctuation, its value is as low as 0.74 at doping density about $9\times10^{16} \mathrm{cm^{-3}}$, at which doping density there is also a kink. The conversion efficiency has a large increase compared to the result in Fig.~\ref{fig:fig4}(d), its maximum value is 21.6\% when there is no doping, it decreases to the smallest value 18.8\% at about doping density $8\times10^{16} \mathrm{cm^{-3}}$, which is equal to $N_I$, then increases to 20\% as doping density continues to increase.

Comparing the results above, we can draw the conclusion that when the doping density is large, such as doping density greater than $1\times10^{17}~\mathrm{cm^{-3}}$, the increase of conversion efficiency is mainly attributed to the doping effect on the control GaAs solar cell without QDs. This is due to the IB is fully filled with electrons for the high doping density, and there are no empty states to accept electrons for tansition from the VB to the IB, and the role of QDs are suppressed.

For the QD IBSC with high generation coefficients, Fig.~\ref{fig:fig6}(a) and Fig.~\ref{fig:fig6}(b) show $n$ and $p$ distribution in the host material GaAs for short circuit and output voltage 1~V, respectively, and the electron occupation factor $f$ in the IB is depicted in Fig.~\ref{fig:fig7}. When there is no doping, $f$ is small on short circuit condition, $-fN_I$ is small in Eq.~\ref{eq:poisson}, so the small amount of electrons in the IB have a little coulomb effect on $n$ and $p$. When the output voltage is 1~V without doping, $f$ is about 0.41, the electrons in the IB has a large coulomb effect, and lead to a concave in $n$ and a convex in $p$ as seen in Fig.~\ref{fig:fig6}(b). When doping density increases, under short circuit $n$ in the region between the QD layer and n-region increases evidently, for the output voltage 1~V, $n$ increases and $p$ decreases at the position of the QD layer, and there is a convex in $n$ and concave in $p$ for large doping density.

\subsection{Analysis and discussion}
It is easy to understand the change of $J_{sc}$ and $V_{oc}$ as doping density increases by simply analyzing $G_I$. As shown in Fig.~\ref{fig:fig6}, $n$ and $p$ change as the QD layer's position, the doping density $N_{dope}$ and output voltage $V_{op}$. According to Eq.~\ref{eq:gi}, $G_I$ is a function of $n$ and $p$, considering the QD layer is fixed in the middle of the i-region in our study, so $G_I$ can be written as $G_I(N_{dope},V_{op})$.

\begin{figure}[ht]
 \centering
 \includegraphics[width=0.9\columnwidth,clip]{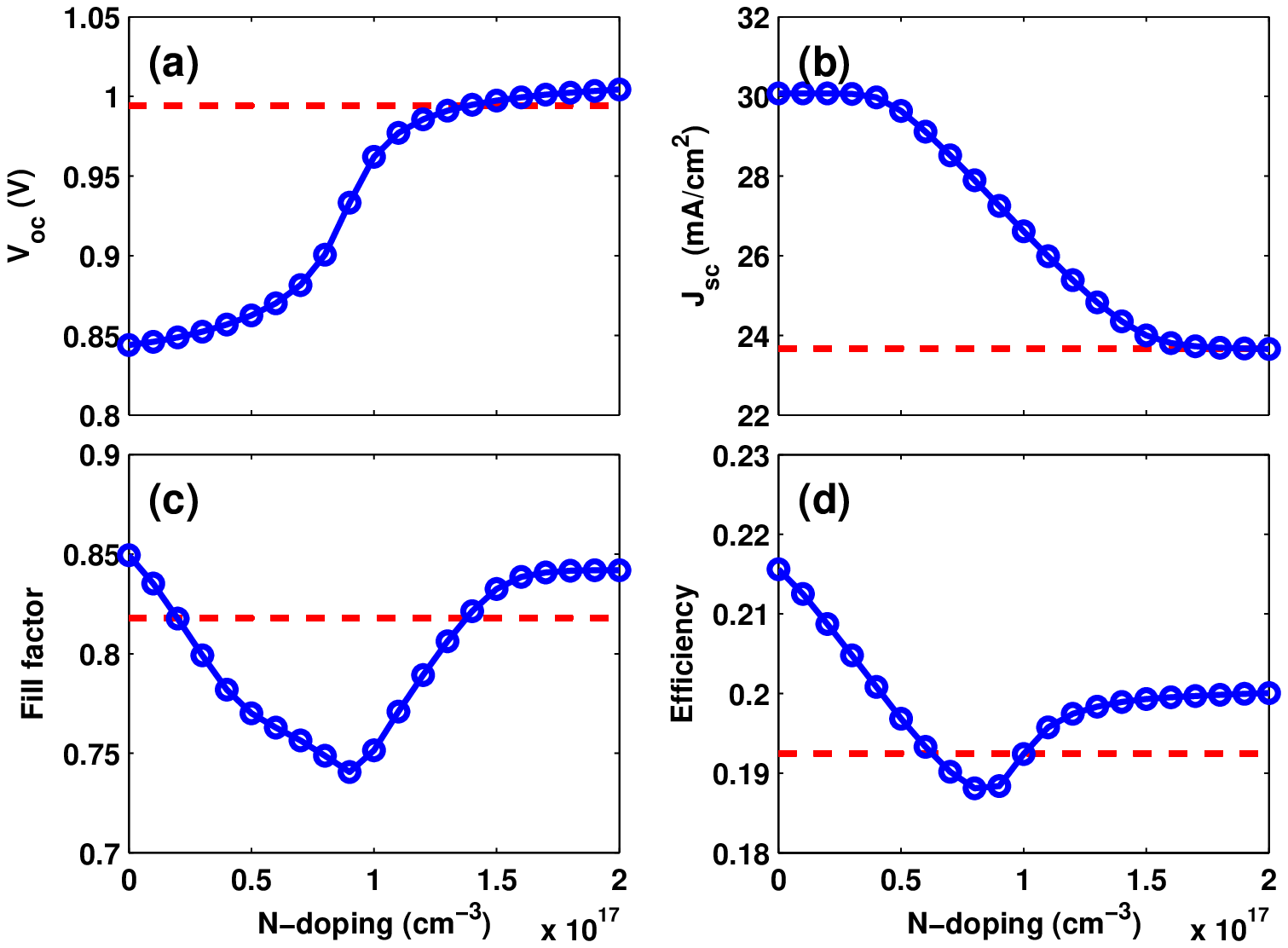}
 \caption{(Color online) For $g_{CI}=g_{VI}=1\times10^5~\mathrm{s^{-1}}$, the current-voltage characteristic changes as the doping density increases. (a) Open circuit voltage. (b) Short circuit current density. (c) Fill factor. (d) Conversion efficiency. Data for the control GaAs solar cell without QD layer are plotted as dashed lines.}
 \label{fig:fig5}
\end{figure}

\begin{figure}[ht]
 \centering
 \includegraphics[width=0.7\columnwidth,clip]{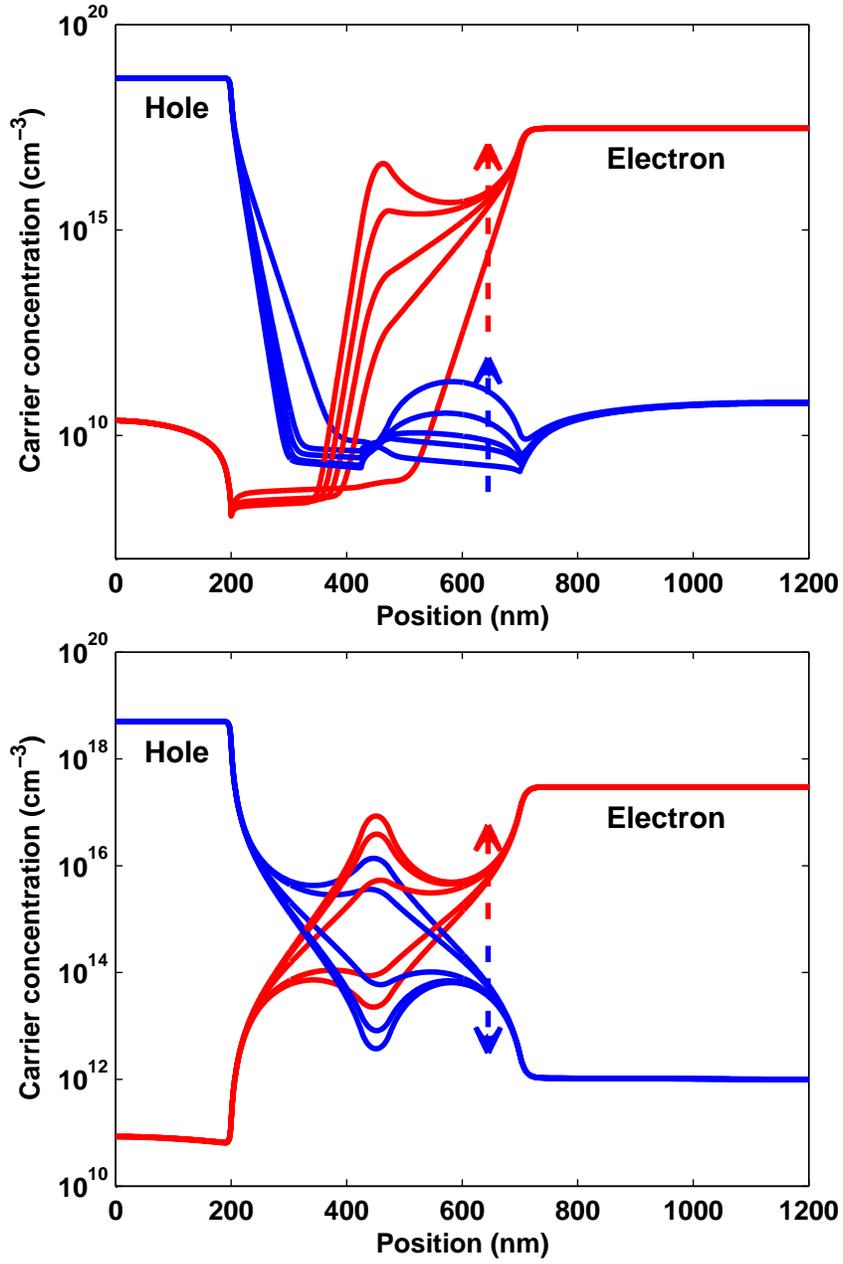}
 \caption{(Color online) For $g_{CI}=g_{VI}=1\times10^5~\mathrm{s^{-1}}$, the carrier concentration of electron and hole in the GaAs host material for the doping density $0$, $5\times10^{16}$, $1\times10^{17}$, $1.5\times10^{17}$ and $2\times10^{17}~\mathrm{cm^{-3}}$. (a) On short circuit condition (b) When output voltage is 1~V. The dashed arrow indicates the change direction as doping density increases.}
 \label{fig:fig6}
\end{figure}

\begin{figure}[ht]
 \centering
 \includegraphics[width=0.7\columnwidth,clip]{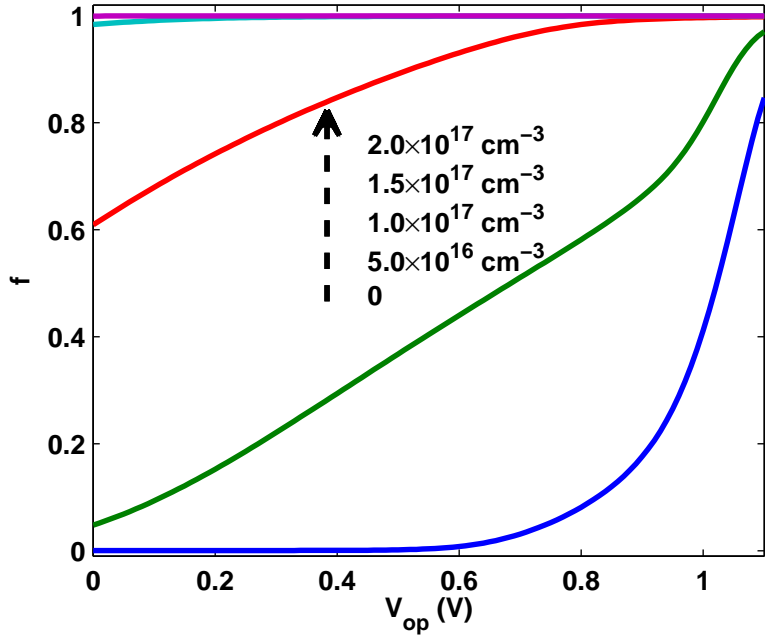}
 \caption{(Color online) For $g_{CI}=g_{VI}=1\times10^5~\mathrm{s^{-1}}$, the electron occupation factor $f$ in the IB vs. output voltage $V_{op}$ with different doping density. The dashed arrow indicates the change direction as doping density increases.}
 \label{fig:fig7}
\end{figure}

On short circuit condition, if there is no doping, the electron and hole carrier concentration are relatively small in the middle of the i-region as shown in Fig.~\ref{fig:fig6}(a), so we can neglect the items about $n$ and $p$ in Eq.~(\ref{eq:gi}), therefore,
\begin{equation}
G_I(N_{dope}=0, V_{op}=0)\approx\frac{N_I(e_eg_{VI}+e_hg_{CI}+g_{CI}g_{VI})}{e_e+e_h+g_{CI}+g_{VI}}.
\end{equation}
Because $e_e\gg e_h$, $e_e\gg g_{CI}$ and $e_e\gg g_{VI}$, $G_I(N_{dope}=0, V_{op}=0)$ can be further simplified, $G_I(N_{dope}=0, V_{op}=0)\approx N_Ig_{VI}$, this means $G_I$ is fully determined by $g_{VI}$ on short circuit condition without doping, compared to the control GaAs solar cell its increase $\Delta J_{sc}$ is $0.15~\mathrm{mA~cm^{-2}}$ in Fig.~\ref{fig:fig4}(b) and $6.4~\mathrm{mA~cm^{-2}}$ in Fig.~\ref{fig:fig5}(b), fully conformed to the relation
\begin{equation}
\Delta J_{sc}\approx eG_Iw_{QD}=eN_Ig_{VI}w_{QD},\label{eq:deltaj}
\end{equation}
where $w_{QD}$ is the thickness of the QD layer 50~nm. When considering the doping, $n$ at the position of the QD layer increases, so $n$ can not be neglect,
\begin{equation}
G_I(N_{dope}, V_{op}=0)\approx\frac{N_I(e_eg_{VI}-r_{CI}r_{VI}np)}{e_e+r_{CI}n}.
\end{equation}
Apparently, $G_I$ decreases monotonically as the doping density increases, so $J_{sc}$ decreases monotonically.

On open circuit condition, according to Fig.~\ref{fig:fig6}(b), $n$ and $p$ are large, considering $e_e\gg e_h$, $e_e\gg g_{CI}$ and $e_e\gg g_{VI}$, so 
\begin{equation}
G_I(N_{dope},V_{op}=V_{oc})\approx-\frac{N_Ir_{CI}r_{VI}pn}{e_e+r_{CI}n+r_{VI}p}.
\end{equation}
Due to the increase of $n$ and $r_{CI}\gg r_{VI}$, $|G_I|$ gets smaller, that is the recombination via IB is reduced, so $V_{oc}$ increases. 

By now literatures all point out that the IB should be half-filled.\cite{marti_partial_2001, luque_operation_2006, strandberg_photofilling_2009, lin_model_2009} But in our simulation model, there is no such relation between the conversion efficiency in Fig.~\ref{fig:fig5}(d) and its corresponding electron occupation factor in Fig.~\ref{fig:fig7}. With the increase of the doping, the predicted increase of light current does't appear in our model. Because the maximum short circuit current density is determined by Eq.~\ref{eq:deltaj}, and can't be increased more by doping. Additionally, by comparing Eqs.~\ref{eq:f0_2} and \ref{eq:f}, although the IB can be half-filled by doping, $f_0=0.5$, $f$ is no long 0.5 when the solar cell works at the maximum power poit.

It is noteworthy that the conversion efficiency are also affected by the fill factor, in Fig.~\ref{fig:fig5} when the doping density changes from 0 to $5\times10^{16}~\mathrm{cm^{-3}}$, the short circuit current density almost has no change, but the conversion efficiency decreases due to the decrease of the fill factor.

\section{Conclusions}
In conclusion, the current-voltage characteristic of the IBSC is affected by the n-type doping density in the QD layer. The open circuit voltage, short current density, fill factor and conversion efficiency all vary with the doping density. As the n-type doping density increases, the open circuit voltage increases, while the short circuit current density decreases, and the conversion efficiency tends to close to that of the control solar cell. In one word, the n-type doping tends to suppress the role of QDs, whether the QDs originally play positive or negative role. This prediction has been observed in our recent experiment, the details about growth, fabrication and experimental data will be published later.

\begin{acknowledgments}
This work is supported by the National Natural Science Foundation of China (Nos.60876033, 61076050, 61021003 and 61176047) and the National Basic Research Program of China (No. 2012CB932701).
\end{acknowledgments}

\newpage

\newpage
\begin{table}[ht]
\caption{\label{tab:parameters} Simulation parameters}
\begin{ruledtabular}
\begin{tabular}{lr}
Energy gap of GaAs $E_g$~(eV)                                               & 1.424  \\
IB energy level $E_I$~(eV)                                                  & 1.124  \\
Permittivity $\varepsilon~(\varepsilon_0)$                                  & $12.9$     \\
Intrinsic concentration of GaAs $n_i~\mathrm{(cm^{-3})}$                    & $2.25\times10^6$  \\
Density of states of the IB $N_I~\mathrm{(cm^{-3})}$                        & $8\times10^{16}$ \\
Donor doping density $N_D~(\mathrm{cm^{-3}})$                               & $3\times10^{17}$ \\
Acceptor doping density $N_A~(\mathrm{cm^{-3}})$                            & $5\times10^{18}$ \\
SRH electron lifetime for GaAs $\tau_n~(\mathrm{s})$                        & $2\times10^{-9}$ \\
SRH hole lifetime for GaAs $\tau_p~(\mathrm{s})$                            & $4\times10^{-7}$ \\
CB to IB recombination coefficient $r_{CI}~(\mathrm{cm^3~s^{-1}})$          & $1.25\times10^{-5}$   \\
IB to VB recombination coefficient $r_{VI}~(\mathrm{cm^3~s^{-1}})$          & $1.25\times10^{-8}$ \\
Surface recombination coefficient $s_{nL}, s_{pR}~(\mathrm{cm~s^{-1}})$     & $1\times10^4$\\
Surface recombination coefficient $s_{nR}, s_{pL}~(\mathrm{cm~s^{-1}})$     & $1\times10^7$\\
CB to VB direct recombination coefficient $r_{CV}~(\mathrm{cm^3~s^{-1}})$   & $7.2\times10^{-10}$  \\
\end{tabular}
\end{ruledtabular}
\end{table}

\newpage

\newpage

\newpage

\end{document}